# WIRELESS SENSOR NETWORKS SIMULATORS AND TESTBEDS

Souhila Silmi[1,2], Zouina Doukha[1], Rebiha Kemcha[2,3]
and Samira Moussaoui[1]

[1]Department of Computer Science, USTHB University, RIMAA Lab., B P N°32 El Alia, 16000 Bab Ezzouar, Algiers, Algeria
[2]Department of Computer Science, Higher Normal School Elbachir El-ibrahimi-Kouba, B P N°92 16308 Vieux-Kouba, Algiers, Algeria
[3]Department of Computer Science, University of Boumerdes, LIMOSE Laboratory, Boumerdes, Algeria

*ABSTRACT*

*Wireless sensor networks (WSNs) have emerged as one of the most promising technologies for the current era. Researchers have studied them for several years ago, but more work still needed to be made since open opportunities to integrate new technologies are added to this field. One challenging task is WSN deployment. Yet, this is done by real deployment with testbeds platforms or by simulation tools when real deployment could be costly and time-consuming. In this paper, we review the implementation and evaluation process in WSNs. We then describe relevant testbeds and simulation tools, and their features. Lastly, we conduct an experimentation study using these testbeds and simulations to highlight their pro and cons. As a use case, we implement a localization protocol. This work gives clarity to future-work for better implementation in order to improve reliability, accuracy and time consumed.*

## 1. INTRODUCTION

The continuous evolution of the need for study and performance evaluation of complex applications, such as security, supervision, military applications, medical and environmental applications, has given importance and necessity to several tools for implementation of Wireless Sensor Networks (WSNs). Recently, WSNs monitor our cities and living environments [1]. With the advent of new technologies, WSNs are integrated with other emerging technologies, giving birth to new architectures such as Vehicular Ad-hoc Networks, Hybrid Sensor and Vehicular Networks (HSVN) , Internet of Things (IoT), and Smart Cities. However, WSNs are difficult to deploy for many reasons. The complexity of the environment, in which the system operates leads to many challenging issues for the designers [1]. The need for adequacy study of the proposed solutions to the reality motivated the development of various tools of test and implementation for new protocols. For that, several tools are available namely: analytical methods, simulations tools, emulation, and prototype generation.  In fact, after the use of these implementation methods, a risk of imperfection towards the reality remains. Actually, it is common to estimate the protocol in a real and concrete context. In this direction, adapted tools emerge, not only to make it more accessible task, but also to master the deployment of these networks.

In several research institutions, testbeds are proposed for various research experiences. Typically, they consist of sensor nodes deployed in a controlled environment, and provide a platform for experimenting with large projects. Research on WSNs is highly expensive when real-time sensors are deployed in specific environment due to constraints on complex topology, and the area of deployment. Thanks to these testbeds, large-scale of resources becomes available. This challenge requires the implementation of simulation environment, with regard to real conditions. In WSN, the availability of multiple simulation tools, making the choice of researchers more difficult to take. In this study, we are interested in a comparative study between the relevant testbeds and simulators in the literature in view of criteria as processing capabilities, accessibility, types of available tests and reliability of results. To the best of our knowledge, there is no comparative study, in the literature, that includes both simulators and testbeds with specific use case.

This paper is divided into six parts: the second section summarizes the existing studies in the field of WSNs implementation. Section 3 presents the simulation tools, their characteristics and practical remarks about the studied simulators. Then, in section 4, we describe the different testbeds in WSNs where we give our remarks for each testbeds studied. Section 5 defines the localization function in WSNs as a use case for our experiments and gives our results with different approaches of implementation. Finally, in section 6, we conclude and present our research open issues that need to be investigated in future work.

## 2. RELATED WORK

The protocols engineering implies numerous phases of tests to validate new solutions. In the specific domain of WSNs, experiences are often complex, hard to repeat and slow to configure and execute. For these reasons, the simulation was the methodology widely used by researchers. However, the researchers become more and more conscious of the fact that the current simulators are unable to model certain essential characteristics of the real systems. For this reason, and due to a visible degradation of specific standards in the driving simulation studies, the simulation results are often debatable and subject of credibility [2]. Therefore, we are aware of a fundamental importance to take forward the theoretical design. The analysis of protocols and algorithms serve in a parallel to the experimental validation, by the use of simulators tools. For such a purpose, the testbeds experiments are proposed.

Several surveys exist in the literature on different research objectives. Some of them provide a comparative study between testbeds. In [3], supporting standards, storage and physical architecture for three platforms are given. A resume about some simulators, with a comparative table about the five highly evaluated tools is given in [4]. In order to explore synergy, authors in [5] review prominent research projects for the different standards and technologies used by platforms. Before giving some detail about testbed, we can find in [6], general information about how to prepare and exploit experimentations in several fields. This survey proposes divers characteristics of protocols for IoT and WSN. It discusses features such as heterogeneity and scaling for the material aspect. Egea-López and all, in [7], study the basic properties to select the appropriate simulation environment. In [8] a comparative study shows that simulators, with regard to VANETs environment, must revise more characteristics. We find works that propose new testbed/simulator, or new architectures. For instance, the authors of [9] intend to give a description of a flexible and non-specific management system, unlike the current management systems, which are strongly coupled to a specific testbed configuration. After providing a summary of the best-known testbeds in [10], this paper proposes a new data recovery mechanism that aids to store the results of structured data in this testbed. This article gives mainly the software aspects of the work. Others propositions are given in [11]-[13].

This paper addresses the comparison between the most popular simulators for WSNs and a comparison between relevant testbeds. A use case is presented to study the effectiveness of two simulators, and two testbeds.

## 3. THE SIMULATION

The simulation of network is definitely one of the most dominant evaluation methods in the field of networks. It is widely used for the development of new communication architectures and network protocols. A simulator is software that imitates the behavior real world components and is used as a research and development tool. Depending on the intended use of the simulator, different parts of the system are modeled. Previous simulators designed for WSNs model the wireless transmission in the network, but currently sensor network simulators have a more detailed model and realistic, including barriers between nodes [14]. A more abstract model is given by the recent simulation tools to get closer to reals environments.

### 3.1. Necessity of the Simulation

The emergence of WSNs has brought new problems to network designers. Computer simulations, analytical methods, or physical measurement (testbeds) can achieve the evaluation of networks performance. Real experiments are important to network research. Although they use complex coding and laboratory experiments, they are able to give important details automatically. However, this approach has disadvantages; real experiments are more expensive to build, and adapting the configuration of laboratory scenarios can be difficult. This leads to limited power sharing and flexibility.

Another point, which can make the evaluation and comparison of protocol designs even harder, is the difficulty to reproduce experimentally as some networking phenomena such as wireless radio interference. The algorithms complexity, in WSNs, is related to their constraints such as limitations due to energy, fault tolerance and decentralized collaboration. Therefore, it appears that the simulation approach is the most feasible approach for quantitative analysis of sensor networks before a real deployment [15]. Simulation allows doing the tests at lower cost and making important decisions.

### 3.2. Simulation Tools

#### 3.2.1. Network Simulator (NS-2)

NS-2 [16] is one of the most widely used simulators in research laboratories to simulate and study the performance of network protocols. It was developed using discrete event technique for network research. It started as a variant of the REAL network simulator [17] in 1989 and evolved considerably nowadays [4]. It has a modular approach and the simulations are based on a combination of C ++ and Otcl languages. NS-2 offers the opportunity to test, analyse and evaluate applications before considering the practical implementation in real networks. NS-2 simulator, particularly, is well suited for packet switching networks and large scale. It contains the necessary functionalities for the study of unicast or multicast routing algorithms, transport, session, booking protocols, integrated services and localization protocols. NS-2 use Network Animator (NAM) as graphical visualization tool.

### 3.2.2. NS-3 Simulator

The NS-3 project, started in 2006, like an open-source project developing NS-3. The NS-3 Simulator is a discrete-event network simulator oriented mainly for research and educational use [18]. Unlike its predecessor NS-2, NS-3 Simulator is based only on C ++ language for the implementation of simulation models. Thus avoiding problems caused by the combination C ++ and oTcl in NS-2. Therefore, network simulations can run on a purely C++ environment, while NS-3 users can possibly run simulations using Python as well [19]. In addition, NS-3 integrates architectural concepts and the GTNetS Simulator code that has good scalability features. The new features appeared with NS3, were made at the expense of compatibility. In fact, NS-2 models have to be brought to NS-3 manually. Several external animators and visualization tools can be used with NS-3 like NS-3-Viz, PyViz and NetAnim [18].

### 3.2.3. Tossim Simulator

Tossim is a simulator for TinyOs operating system, and it was developed at UC Berkeley [20]. It simulates the behavior of a sensor (sending/receiving messages via radio waves and processing information). This simulator is written in NesC Language which provides a component-base programming model. TOSSIM simulates the TinyOs network stack at bit level, allowing experimentation with low-level protocols in addition to high-level application systems. TOSSIM is based on the assumption that each node in the network must run exactly the same code, which makes it less flexible. TOSSIM does not model energy consumption, so PowerTOSSIM Simulator and PowerTOSSIM z are an improvement that extend the simulator to model energy consumption [21].

### 3.2.4. OMNet++ Simulator

Unlike ns-2 and ns-3, OMNet ++ is not a network simulator by definition, but a discrete framework of general use based on simulation events. Although it is most often applied to the field of network simulation, given that its INET package offers a complete collection of Internet Protocol models. In addition, it has other models packages like the mobility package for mobile Ad-Hoc networks and mobile WSNs [18].

### 3.2.5. Avrora Simulator

It is an open source simulator for embedded detection programs. The current version (version 1.7.106) is written in Java. It models two typical platforms, Mica2 and MicaZ [22]. It also provides a framework for program analysis, static verification of embedded software and infrastructure for future programs. Avrora simulator is flexible providing a JAVA API to develop analytics [23] [24].

Table 1. Comparative table of simulation tools in WSNs.

| Simulators | Language | Main characteristics | Limits |
|---|---|---|---|
| NS-2 | C++ / OTcl | + Easy to add new protocols<br>+ A large number of publicly available protocols<br>+ Has a visualization tool<br>+ Has a rich collection of models | - Designed for wired networks basically<br>- Uses pure source code in terminal for the development of the simulation set |
| NS-3 | C++ | + Perform large-scale network simulations efficiently [18]<br>+ Low calculation and memory requests<br>+ Has a visualization tool | -Difficult to implement and simulate a protocol than NS-2 [18] |
| TOSSIM | NesC (dialect of C) | + High degree of accuracy by using a few low-level components models [24]<br>+ Has a visualization tool | - Simulates only TinyOS programs [20]<br>- Can only emulate the homogeneous applications for the same type of sensors |
| OMNet++ | C++ | + Perform large-scale network simulations efficiently<br>+ Has a visualization tool<br>+ Has an abstract modelling language | - Unavailability of several important protocols<br>- The combination of models may be difficult and programs are more likely to have bugs [18] |
| AVRORA | Java | + An instruction-level sensor network simulator<br>+ Has TOSSIM performance while preserving cycle accuracy [24]<br>+ Portable due to use of Java virtual machine | - Does not have network communication tools<br>- Does not have GUI<br>- Cannot simulate network management algorithms [24] |

### 3.3. From Simulation Tools to Experimental Testbeds in WSNs

The study shows that each simulator has strengths and limitations, as shown in Table 1. However, the challenge of development, deployment and debugging of applications in realistic environment may not be satisfied with simulation. Many current simulators are unable to model several essential features of the real world. Simulation results are only as good as the model then are only estimated results. Especially for simulation models of WSNs which do not capture the irregularity in the network, either for radio irregularity on the communication [25]; or the irregular sensor model [26], for example. Recent material spread has allowed for new uses for protocol designers. Actually, after the classical phases of specification and simulation, it is common to evaluate a protocol in a real and concrete context. So, adaptive tools are emerging, not only to make this task more accessible, but also to master its deployment, in order to have more usable results.

## 4. TESTBEDS FOR WIRELESS SENSOR NETWORKS

During our research, we found several testbeds. We have made a choice for the most significant ones and we highlight their characteristics to make it easy for researcher's investigations.

### 4.1. W-iLab.t

W-iLab.t is a Wireless testbed that enables to test wireless protocols or applications in a real-time environment [27]. The IBBT iLab.t technology Centre manages this testbed. It is based on MoteLab testbed, and is located in Ghent, Belgium. W-iLab.t testbed has a web interface to plan, download, monitor the experience and recover the results. Only authorized users can access the interface. To enable mixed Wi-Fi node and sensor node experiments and to keep a uniform interface, they integrate the support for the Wi-Fi nodes into the same web interface as used for the sensor nodes [28]. Unique features of the test include real-time monitoring of energy consumption and emulation of battery capacity. W-iLab.t has various types of wireless nodes, which are connected to a wired interface for management, and then they can be used during experiments as a cable interface. Heterogeneous experiments wireless/wired are possible. Experimenters are able to give the complete configuration for each device in the network. With this testbed, measurement and management data are saved in a database. This data is made available to the user to be exploited either for processing or for only visualization.

### 4.2. Tutornet

Tutornet [29] is a testbed for WSNs at Southern California University. It has three levels; test servers, bridge stations and sensor nodes. This testbed includes 104 sensor nodes (91 TmoteSky and 13 MicaZ). The nodes are attached to the gateway stations via USB connections. A Star-gate with several nodes around it form a cluster. Currently there are 13 clusters and the nodes can be programmed distantly.

### 4.3. MoteLab

MoteLab [30] is a testbed for WSNs at Harvard University. It is accessible for the development and testing of sensor network applications via a web interface. Registered users can download and associate executable files to nodes, to create and plan an experiment to run. MoteLab aims to facilitate research in programming sensor network environments, communication protocols, system design and applications. This testbed contains 190 TMote Sky sensor nodes. The motes consist of the MSP430 processor operating at 8MHz, 10KB RAM, 1Mbit flash memory and a Chipcon CC2420 radio operating at 2.4GHz with an internal range of 100 meters. Light, temperature and humidity sensors are integrated in each node. Nodes run the TinyOS operating system and use the NesC language.

### 4.4. Wisebed

Wisebed [31] is an experimentation platform in WSN for research purposes. This platform is distributed on nine geographically dispersed sites; it makes available to users more than 750 heterogeneous sensors. Once the registration is confirmed, the user has access to start the configuration of the nodes, as it is depicted in Figure 1.

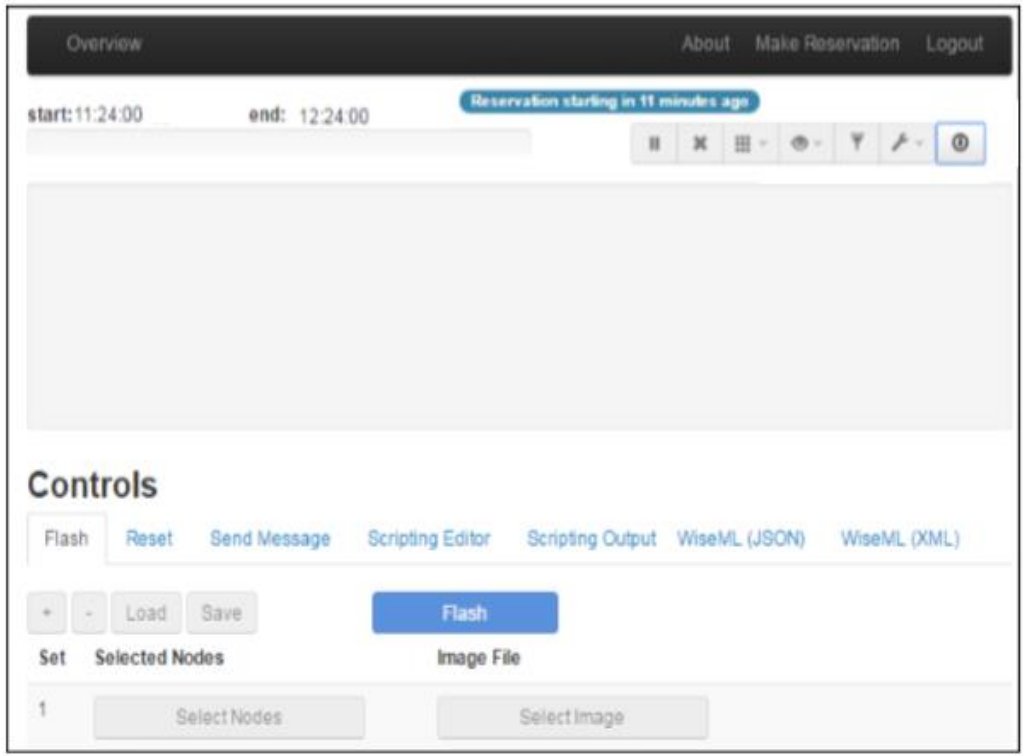

Figure 1. Experiment in WISEBED testbed.

## 4.5. FIT IoT-LAB

FIT IoT-LAB (IoT-LAB) is a proposed testbed to assure a large infrastructure for scientists in WSNs field. Heterogeneous communicating objects and small wireless sensor devices can be easily tested by this platform. It is an improvement of the SENSLAB testbed (2010-2013) [32], which is located at six different sites in France, with 2728 heterogeneous nodes. IoT-LAB testbed is a well designed system so that user can specify needed properties like the location, radio chip and whether nodes are mobile or not. It allows the repeatability of experiments [6].

## 4.6. Nitos

The Network Implementation Testbed (NITOS) is developed by NITLAB (Network Implementation Testbed Laboratory) using open source platforms at University of Thessaly in Greece. The NITOS installation currently includes more than 100 operational wireless nodes and is designed to ensure reproducibility of experimentation while supporting the evaluation of protocols and applications in real-world environments. This platform offers testbeds for several types of networks technology [33]. The NITOS testbed includes three different testbeds: the NITOS outdoor testbed, the NITOS indoor testbed, and the NITOS office testbed. All these testbeds are connected with a backbone connection provided by the Greek NREN. They use a part of the pan-European GEANT network [34]. The NITOS testbed is open to all researchers wishing to test their protocols in real-world environments. They have the opportunity to implement their new protocols and study their behavior in a parameter-able environment.

### 4.7. Planet Lab

Planet Lab [35] is an open platform, shared on a large scale to conduct real world experiments to develop new network services and technologies. It is developed in March 2002 by a network of international private and public laboratories [36]. Although Planet Lab is not dedicated to sensor networks, it is associated with different types of testbeds, including testing on WSNs as part of the initiative to develop testbeds for future Internet. This platform comprises currently more than 1000 nodes on 500 sites throughout the world including Asia, USA, Europe, and Japan. There is also OneLab testbed initiative, which extends Planet Lab Europe by uniting it with other Planet Lab testbeds around the world as well as other types of tests.

### 4.8. ORBIT

The ORBIT project provides a flexible wireless network testbed open to the experimental researchers [37]. It is designed to achieve reproducible experimentation while supporting realistic evaluation of protocols and applications. ORBIT was funded in 2003 under the Network Research Testbeds (NRT) program [11]. Figure 2 shows the ORBIT radio grid, which was first released to researchers. Currently, it is heavily used when evaluating architectures and protocols in wireless networks. In this testbed, tests on nodes-sensors is given via the Linux terminal [38].

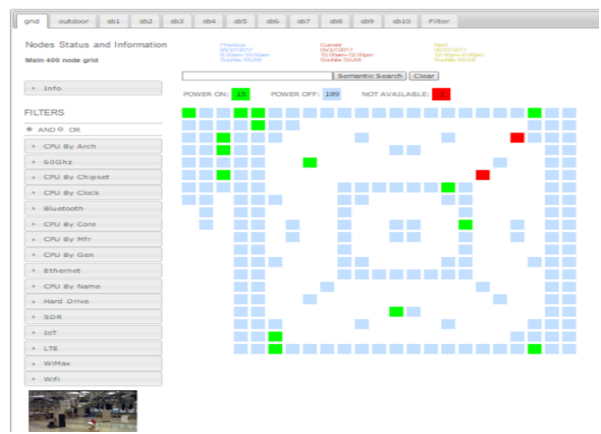

Figure 2. Radio Grid testbed for ORBIT

### 4.9. Twist

TWIST [39] is an evolutionary and flexible testbed, which is developed by the Telecommunications Network Group (TKN) at Technical University Berlin. It supports heterogeneous hardware and deploys 204 nodes (102 TmoteSky nodes and 102 eyesIFXv2 nodes). Twist testbed allows testing algorithms on different platforms and provides basic services such as node configuration, network programming, off-band retrieval of debugging data and application data collection, as well as several specific functionalities. The lowest level of its architecture is connected by USB cabling and USB hubs to the test infrastructure. The second level contains a «super nodes», which can interact with the USB infrastructure of first level; and the last level has the server and control stations that interact with the super nodes using the testbed backbone. The server stores information on registered nodes in its database and provides remote access via a web interface as depicted in Figure 3. This testbed supports different communication protocols such as Wi-Fi, Zigbee and Bluetooth. Tests on the reserved sensors can be done either via the graphical interface (called Jfed interface) (like for W-ilab.t2 testbed), or via the Linux terminal (like for Orbit testbed).

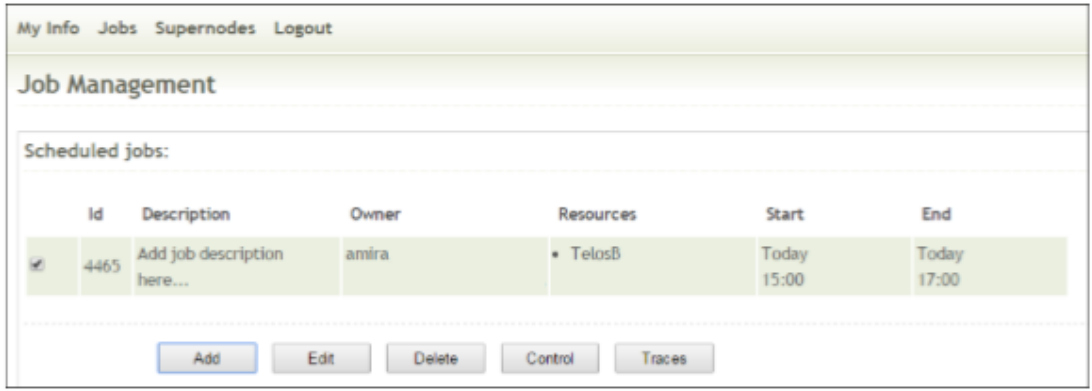

Figure 3. Twist reservation of ressource manager

### 4.10. CERIST

CERIST testbed is developed by the Research Center for Scientific and Technical Information in Algeria. It has more than 30 sensor nodes (10 Telosb and 21 MicaZ). Figure 4 shows the testbed environment.

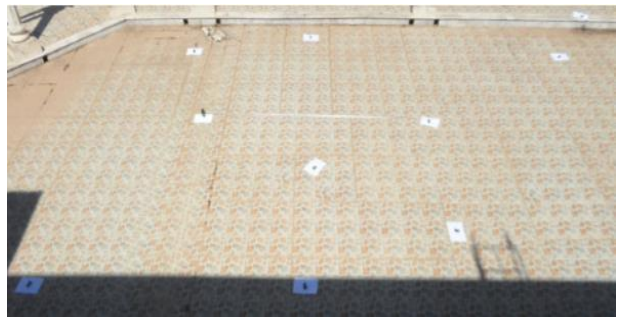

Figure 4. Sensors positioning in CERIST testbed.

Table 2.  Comparative table for testbeds in WSNs.

| Testbed | Authorized Population | OS | Sensor type | Advantages | Limits |
|---|---|---|---|---|---|
| W-ilab.t | Open access | TinyOS | Zolertia Z1 | + Graphic interface. <br> + Accessible via EmuLab portal [40] | - Competition of reservation. <br> -Request to the authority IMinds |
| TutorNet | Open access | TinyOS | 91 Tmote Sky 13 MicaZ |  | - Incorrect site link |
| MoteLab | Open access | TinyOS | TMote Sky | + Graphic interface. <br> +Uses database | -No Web portal. <br> - Incorrect site link |
| NITOS | Open access | Contiki | M3/A8 tiny NITOS mote | +Heterogenous testbed. <br> +Accessible via OneLab portal | - Competition of reservation. |
| IoT-LAB | For members | Contiki FreeRtos | Msn430, M3,M8 | +Many platforms. <br> +Accessible via OneLab portal. | - Director's agreement is required for |

| Testbed | Authorized Population | OS | Sensor type | Advantages | Limits |
|---|---|---|---|---|---|
| | | TinyOs Riot, OpenWsn | | | students |
| Orbit | For members | TinyOS | Tmotes, Telosb | +Comfortable interface. | -Not scalable -Competition for reservation. |
| PlanetLab | For members | TinyOS | | + Compatible applications with Linux and Fedora8 + Accessible via OneLab portal. | - Competition of reservation. |
| Twist | Open access | TinyOS | Tmote Sky eyesIFXv2 | +Experimentation on different platforms. +Heterogenous testbed. | - Does not detect automatically the connected notes [39] |
| CERIST | Open access | TinyOS | Micaz / Telosb | + Personalisable environment | -No web portal -Not scalable |

## 4.11. Discussion

Although testbeds are diverse and their services are rich and attractive, they are still subject to credibility. Table 2 summarizes our results and remarks about relevant studied testbeds.

As remote interaction with resources is always dependent on the reliability of internet connection, several times, we needed to redo the whole procedure of preparing the environment and test files due to the interruption of the connection to the web interface.

Account creation seems easy and instantaneous, but validation access can last longer. Most testbeds require selecting the organization to which the user belongs; otherwise, access is not possible for foreign students (Tutornet, OneLab and MoteLab cases).

Sometimes, the resource reservation system encounters some access contention, and this can cause the interruption or the cessation of experiments (ORBIT case).

Maintenance of websites without prevention (WISEBED case) and the greater level of access concurrence makes the available hours too limited, and it may be impossible to complete all tests (W-iLab.t case, when our reservation is not achieved due to unavailability of service with maintenance works). Further, the exhaustion of time reservation implies the immediate disconnection of the server, which can cause data loss and discontinuation of tests.

Most testbeds provide web interfaces for task planning (FIT IoT-LAB and WISEBED cases).

To manage requests, IoT-LAB and WISEBED testbeds use a first come first serve service, which means that the first user requesting experience on available resources gets access.

Testbeds tutorials are not comprehensive enough and explanatory (TWIST case, when access to the sensors in finale step requires a password that was provided to us by the administrators and TWIST tutorials do not mention it).

## 5. PROTOCOL IMPLEMENTATION IN WSNS

### 5.1. Localization Protocol as a Use Case

Among the fundamental functionality studied in WSNs, one is the localization of sensors. In the literature, several systems and algorithms deal with this problem. However, research studies aim to develop cost-effective and self-localization algorithms in order to improve the performances. The existing solutions can be classified according to the adopted approach as twofold: The fine localization and the approximate localization. The fine localization approaches accurately determine the coordinates of each node while coarse localization approaches give non-precise position. For the approximate localization approaches, node will use other techniques to estimate an approximate localization. In this section, we study the performances of existing protocols, with regard to simulation and testbed implementation, to determine the gap between the obtained results and see how much the results are reliable.

Heurtefeux and Valois in [41] propose a Qualitative Location Protocol (QLoP). It is considered as one of the most powerful protocols that gives approximate localization for sensor nodes. They use local connectivity information to allow every node to determine the proximity coefficient of its neighbors in one hop [42]. QLoP protocol considers a qualitative metric to describe neighbors like very close neighbor, close or far.

### 5.2. Implementation with Simulation

In QLoP Protocol evaluation work, Avrora and NS2 simulators conducted our simulation tests.

#### 5.2.1. Operating System and Programing Language

There are several operating systems for sensor networks such as TinyOS, Contiki, Mantis OS, Nut/OS and SOS. TinyOs is the most famous and most complete for several types of sensors like the Mica family, the Telos family and the Iris family. In our simulations, we used TinyOs sytem with NesC language. NesC language has a component-based architecture. The implementation of components is done by declaring tasks, commands or events.

#### 5.2.2. Simulation Environment Parameters and Metrics

The static nodes were placed regularly on a grid in a surface of 100 * 100 meters. Table 3 shows our simulation parameters, as the transmission range (R) is used to control the average degree of nodes in the network. The number of well-located nodes is calculated as one of the metrics, which estimate the accuracy.

Table 3. Simulation parameters.

| Paramter | Value |
| --- | --- |
| Nodes number ( N ) | 100 |
| Communication range ( R ) | 10, 15, 20 meters |
| Sensor type | MicaZ |
| Topology | Static |
| Simulator | Avrora, NS2 |

### 5.2.3. Simulation Results

Figure 5 shows the simulation results by applying different communication range. The green nodes, in the figure, represent nodes that classify correctly their neighbors, while black nodes do not classify correctly their neighbors. For nodes on the edges of the simulation space, the algorithms cannot order correctly the neighbors of the first two proximity classes, due to inconsistencies in the neighborhood.

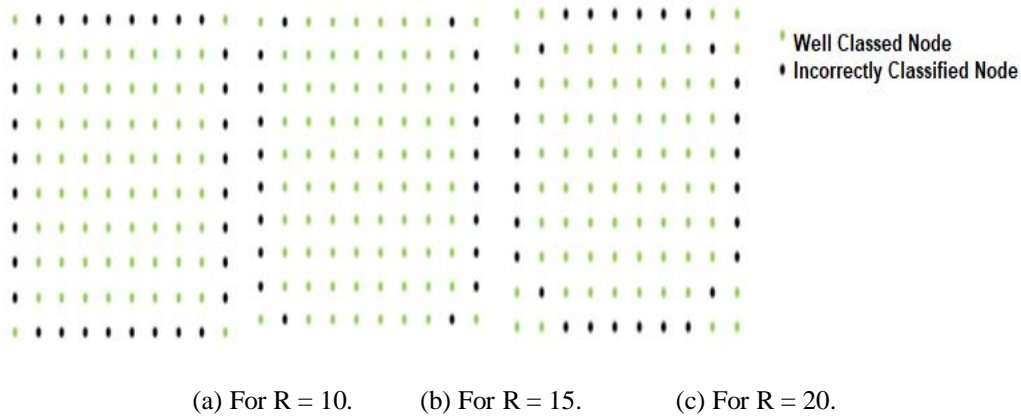

(a) For R = 10.    (b) For R = 15.    (c) For R = 20.

Figure 5. Simulation Results.

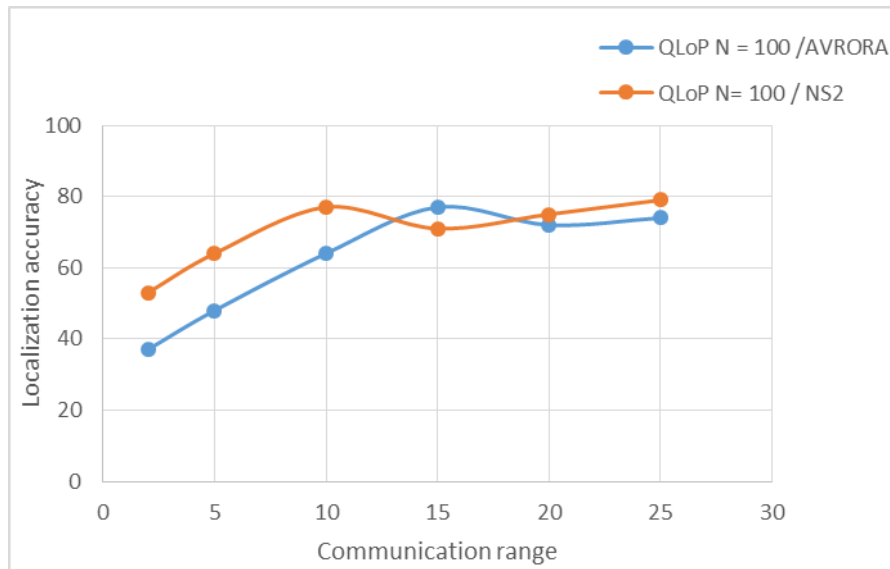

Figure 6. Localization accuracy according to communication range
with NS2 Simulator and AVRORA Simulator.

Figure 6 shows the localization accuracy with NS2 Simulator and AVRORA Simulator. The two curves have the same general speed, although NS2 results for communication range less than 5, starts from an accuracy around 50, contrary to AVRORA results which begins with an accuracy less than 40. In [41], authors prouve with simulation that the localization accuracy is more than 80 for differents communicaion range. Then, NS2 Simulator gave more closed results than AVRORA Simulator.

### 5.3. Implementation with CERIST Testbed

In our real experimentations, we used CERIST testbed. Initially, this testbed worked with Emulab testbed. However, the access to these resources is interrupted [43]. We visited this research center to use the remaining hardware of this platform. We implemented the protocol using two types of sensors: Micaz and Telosb sensors, in collaboration with sensor team of the Center. In our implementation, we used 10 MicaZ and 5 Telosb, as shown in Figure 7.

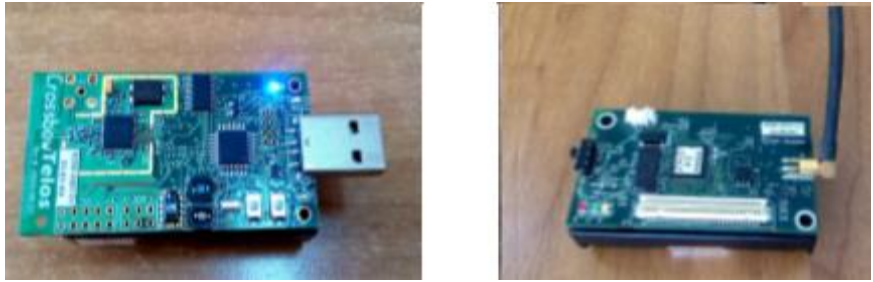

Figure 7. CERIST testbed sensors.

The platform is set up in two parts: software installation and hardware installation. For the software part, we used NesC language with TinyOs system, in Linux environment (Ubuntu distribution). It is possible to use windows environment but there are problems with the second version of Tinyos (Tinyos-2-x) in this environment. We follow many steps to install Tinyos2.1.2, Nesc and mico controler MSP430 for the Telosb and Micaz sensors.

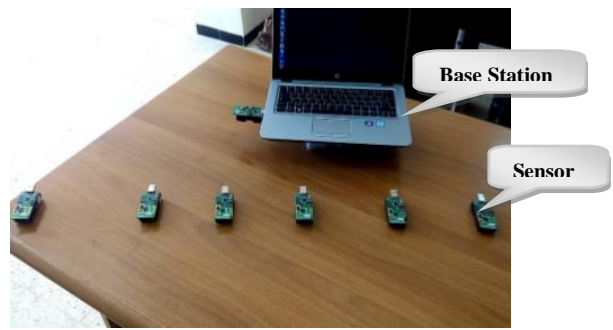

Figure 8. Platform Components.

For the hardware installation, we used a base station connected to the computer via USB, and different sensors like sender / receiver, as shown in Figure 8. Each sensor launches the localization project to classify its neighbors. Then, it sends out the results and makes them available offline in the base station. All sensors communicate via a wireless connection and the base station communicates with the computer via USB.

#### 5.3.1. Implementation Topologies

A network is a set of inter-connected terminals to share information. In our experiments, we used two types of topologies: to reflect reality. The nodes are placed randomly, in the random topology. Nevertheless, the uniform topologies have the advantage of being simple to visualize, then we used also a grid topology. Figure 9 shows an example for each kind of topology.

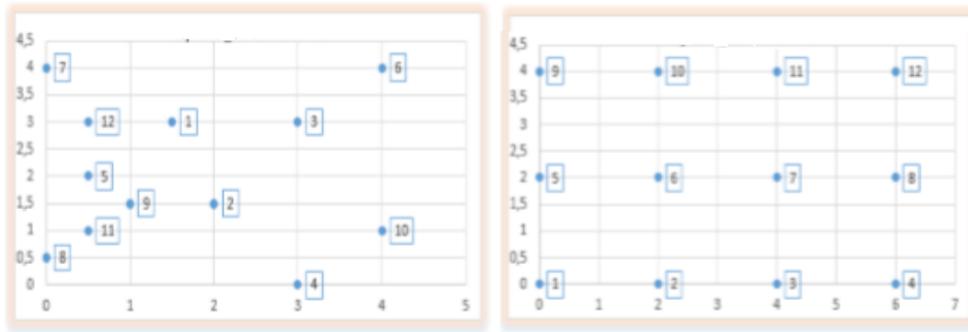

Figure 9. Random Topology and a grid Topology of deployment.

### 5.3.2. Nodes Placement

How to place the nodes in the area of interest is an important factor for the localization problem. Among the placement types, there exist:
-Uniform placement: Example of such placement is the grid. However, it is not suitable for dense network;
-Dense placement: which has cost and signal interference problems;
-Incremental placement: In this type, global calculations are made to estimate the optimal place for each new sensor. This mechanism is recommended for WSNs. The incremental placement needs adaptation after installation to fit the communication requirements;
-Self-deployment placement: It maximizes network coverage and takes into account the
problem of obstacles. The nodes must perform self-deployment placement autonomously.

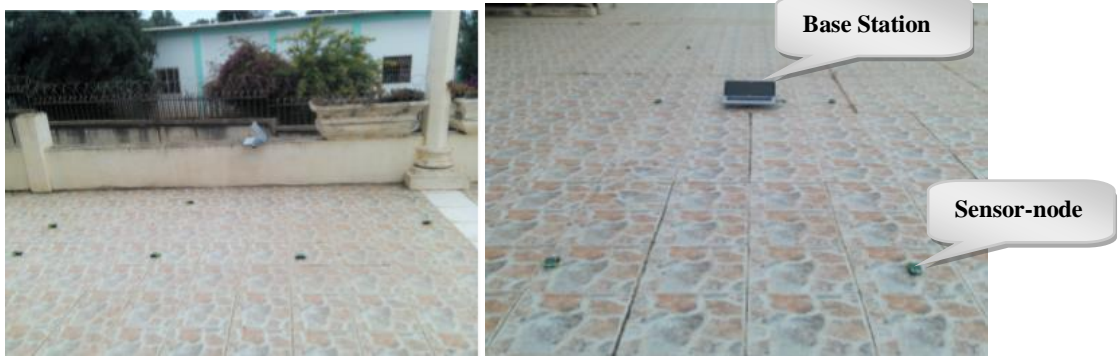

(a) Random Topology.     (b) a grid Topology of deployment.

Figure 10. CERIST testbed demployment.

### 5.3.3. Implementation Results

The network can be implemented with homogeneous or heterogeneous sensor-nodes. With the first, all sensors have the same characteristics (like operating range, sensitivity, finesse). However, the other heterogeneous networks are multi-supplier networks, so the hardware or software components come from different suppliers. In our case, we used 10 Telosb and 5 Micaz, like shown in Figure 10, for a random topology or a grid topology. We used two topologies with the heterogeneous sensor-nodes, and the rests were homogeneous ones.

The metrics used in real implementations are the same used in simulations. The 'Results' column in next tables contains two results: Right and Wrong. 'Right' result represents the node that has correctly classified its neighbor. For example: in table 4, node 1 has correctly classified its neighbor node 4, as its classification is 3 (far class). This classification is equal to the real class. For N = 4, almost all nodes do not correctly locate their neighbors. We have identical results for both topologies.

Table 4. Implementation results for a grid and random topologies for N = 4 nodes.

| Node | Neighbors | Distance (m) | Transmission range (R) | A grid result | Random result | Real class | Note |
|------|-----------|--------------|------------------------|---------------|---------------|------------|-------|
| 1 | 2 | 2    | 3 | 1 | 1 | 2 | Wrong |
| 1 | 3 | 2    | 3 | 3 | 3 | 2 | Wrong |
| 1 | 4 | 2.82 | 3 | 3 | 3 | 3 | Right |
| 2 | 1 | 2    | 3 | 1 | 1 | 2 | Wrong |
| 2 | 3 | 2.82 | 3 | 3 | 3 | 3 | Right |
| 2 | 4 | 2    | 3 | 3 | 3 | 2 | Wrong |
| 3 | 1 | 2    | 3 | 1 | 1 | 2 | Wrong |
| 3 | 2 | 2.82 | 3 | 1 | 1 | 3 | Wrong |
| 4 | 1 | 2.82 | 3 | 1 | 1 | 3 | Wrong |
| 4 | 2 | 2    | 3 | 1 | 1 | 2 | Wrong |

Table 5 and 6 represent the implementation results of QloP protocol for six nodes. For more nodes in the network, we have better results for both topologies. Really, for a random placement of nodes, six nodes could correctly class their neighbors. While in a regular placement, with a grid, we have only four nodes that could correctly class their neighbors. The results were the same for N= 9 and N =12. When the number of nodes is increased, the results improve more and more.

Table 5. Implementation results for a grid topology for N = 6 nodes.

| Node | Neighbors | Distance (m) | Implementation result | Real class | Communication range (R) | Result |
|------|-----------|--------------|------------------------|------------|--------------------------|--------|
| 1 | 2 | 2    | 2 | 2 | 3 | Right |
| 1 | 3 | 2    | 1 | 2 | 3 | Wrong |
| 1 | 4 | 2.82 | 2 | 3 | 3 | Wrong |
| 1 | 5 | 4    | 1 | 3 | 3 | Wrong |
| 2 | 1 | 2    | 1 | 2 | 3 | Wrong |
| 2 | 3 | 2.82 | 3 | 3 | 3 | Right |
| 2 | 4 | 2    | 1 | 2 | 3 | Wrong |
| 2 | 5 | 2.23 | 2 | 3 | 3 | Wrong |
| 3 | 1 | 2    | 1 | 2 | 3 | Wrong |
| 3 | 2 | 2.82 | 2 | 3 | 3 | Wrong |
| 3 | 4 | 2    | 2 | 3 | 3 | Right |
| 3 | 5 | 2    | 1 | 2 | 3 | Wrong |
| 3 | 6 | 2.82 | 3 | 2 | 3 | Right |
| 4 | 1 | 2.82 | 2 | 3 | 3 | Wrong |
| 4 | 2 | 2    | 1 | 3 | 3 | Wrong |
| 4 | 3 | 2    | 2 | 3 | 3 | Right |
| 4 | 5 | 2.82 | 2 | 2 | 3 | Wrong |
| 5 | 1 | 4    | 1 | 3 | 3 | Wrong |
| 5 | 2 | 2.23 | 2 | 3 | 3 | Wrong |
| 5 | 3 | 2    | 1 | 2 | 3 | Wrong |
| 5 | 4 | 2.82 | 2 | 3 | 3 | Wrong |
| 5 | 6 | 2    | 3 | 2 | 3 | Wrong |
| 6 | 3 | 2.82 | 1 | 3 | 3 | Wrong |
| 6 | 5 | 2    | 1 | 2 | 3 | Wrong |

Table 6. Implementation results for random topology for N = 6 nodes.

| Node | Neighbors | Distance (m) | Implementation result | Real class | Communication range (R) | Result |
|---|---|---|---|---|---|---|
| 1 | 2 | 2 | 2 | 2 | 3 | Right |
| 1 | 3 | 2.06 | 1 | 3 | 3 | Wrong |
| 1 | 4 | 1.41 | 2 | 2 | 3 | Right |
| 1 | 5 | 1.11 | 1 | 2 | 3 | Wrong |
| 2 | 1 | 2 | 1 | 2 | 3 | Wrong |
| 2 | 3 | 0.5 | 3 | 1 | 3 | Wrong |
| 2 | 4 | 1.41 | 1 | 2 | 3 | Wrong |
| 2 | 5 | 2.23 | 2 | 3 | 3 | Wrong |
| 3 | 1 | 2.06 | 1 | 3 | 3 | Wrong |
| 3 | 2 | 0.5 | 2 | 1 | 3 | Wrong |
| 3 | 4 | 1.11 | 2 | 2 | 3 | Right |
| 3 | 5 | 3 | 1 | 3 | 3 | Wrong |
| 3 | 6 | 1.11 | 3 | 2 | 3 | Wrong |
| 4 | 1 | 1.41 | 2 | 2 | 3 | Right |
| 4 | 2 | 1.41 | 1 | 2 | 3 | Wrong |
| 4 | 3 | 1.11 | 2 | 2 | 3 | Right |
| 4 | 5 | 1.41 | 2 | 2 | 3 | Right |
| 5 | 1 | 1.11 | 1 | 2 | 3 | Wrong |
| 5 | 2 | 2.23 | 2 | 3 | 3 | Wrong |
| 5 | 3 | 3 | 1 | 3 | 3 | Wrong |
| 5 | 4 | 1.41 | 2 | 2 | 3 | Right |
| 5 | 6 | 1.5 | 3 | 2 | 3 | Wrong |
| 6 | 3 | 1.11 | 1 | 2 | 3 | Wrong |
| 6 | 5 | 1.5 | 1 | 2 | 3 | Wrong |

## 5.4. Discussion

This work is directed as an empirical study of simulators and testbeds in order to allow a better choice in an evolution of a new protocol. It necessarily involves the study of existing simulators and testbeds in the literature, which enables the most used tools of the research community to select the best performers.

Simulation is a perfect environment, but the virtual world of simulator does not completely fit the characteristics of the real world. In fact, the Telosb and Micaz sensors have an omnidirectional antenna by definition. That is, the distance of communication covered by the sensor's transmission row is the same in all directions. Nevertheless, tests have shown that the distance covered (for a given transmission range) is not the same for the four sensor directions. For example: for node 3, the transmission range on one side of the Telosb sensor is 4 meters (m), for another side is 4m30. In addition, Micaz sensors do not work well if placed directly on the ground. It is therefore necessary to find objects like chairs or stones to place them on. Weather conditions have a major impact on communication, such as heat. In fact, the sensors run their program rapidly and efficiently in an indoor environment or in the shade. The number of runs (boot), that we can perform, was limited to 10,000 tries, and the energy of a sensor was limited. During the implementation process with testbeds, we faced some unavoidable situations, which greatly influenced the results of the tests carried out. All these remarks should be taken into consideration to infer results that are more correct.

## 6. CONCLUSION

Wireless sensor networks are widely used in various applications and domains. In this work, we have carried out a detailed experimentation study about implementation technics in WSNs. WSN testbed is designed to support experiments research in the real environments. Currently, experiments can be repeated to give more precise analysis results. In addition, simulation is an important approach for the implementation process, which is very useful for researchers. Although it cannot absolutely replace real experiments with testbeds. However, different communities of researchers can use a standard simulation framework, to increase the reliability and approval of the simulation results.

It is deeply important to offer researchers with this tools, to facilitate design stains, implementation and performance studies. Our study made practical remarks about each approach, because, if simulation gives enthusiastic results, the testbed can give results that are more faithful. It will give options to the researchers in their performances studies. The choice between testbeds and simulators should be based on the specific application as every WSN application has different requirements. These results urge us to develop more tools that include applications requirements in terms of mobility and heterogeneity of technologies, for example.

## REFERENCES

[1]   D. Rong, (2016) "Wireless Sensor Networks in Smart Cities: The Monitoring of Water Distribution Networks Case", (Doctoral dissertation, KTH Royal Institute of Technology).

[2]   S. Kurkowski, T. Camp, & M. Colagrosso, (2005) "MANET simulation studies: the incredibles", ACM SIGMOBILE Mobile Computing and Communications Review, 9(4), 50-61.

[3]   J. YICK, B. MUKHERJEE, D. GHOSAL, (2008)"Wireless sensor network survey", Computer networks, vol. 52, no 12, p. 2292-2330.

[4]   B. I. Bakare, and J. D. Enoch, (2019) "A Review of Simulation Techniques for Some Wireless Communication System", International Journal of Electronics Communication and Computer Engineering, 10(2).

[5]   P. Rawat, K. D. Singh, H. Chaouchi, and J. M. Bonnin, (2014) "Wireless sensor networks: a survey on recent developments and potential synergies", The Journal of supercomputing, 68(1), 1-48.

[6]   A. S. Tonneau, N. Mitton , and J. Vandaele, (2015) "How to choose an experimentation platform for wireless sensor networks? A survey on static and mobile wireless sensor network experimentation facilities", Ad Hoc Networks, 30, 115-127.

[7]   E. Egea-Lopez, J. Vales-Alonso, A. S. Martinez-Sala, P. Pavon-Marino, and J. García-Haro (2005, July) "Simulation tools for wireless sensor networks", In proceedings of the international symposium on performance evaluation of computer and telecommunication systems (SPECTS05) (p. 24).

[8]   S. A. B. Mussa, M. Manaf, K. Z. Ghafoor, and Z. Doukha, (2015, October) "Simulation tools for vehicular ad hoc networks: A comparison study and future perspectives", In 2015 International Conference on Wireless Networks and Mobile Communications (WINCOM) (pp. 1-8). IEEE.

[9]   P. Hurni, M. Anwander, G. Wagenknecht, T. Staub, and T. Braun, (2011, October) "Tarwis: A testbed management architecture for wireless sensor network testbeds", In Proceedings of the 7th International Conference on Network and Services Management (pp. 320-323). International Federation for Information Processing.

[10]  M. Pushpalatha, R. Venkataraman, K. Sornalakshmi, and T. Ramarao, (2013) "Implementation of Wireless Sensor Network Testbed-SRMSenseNet", International Journal of Computer Applications, 63(17).

[11]  E. Ertin, A. Arora, R. Ramnath, V. Naik, S. Bapat, V. Kulathumani, and M. Nesterenko, (2006, April) "Kansei: a testbed for sensing at scale", In Proceedings of the 5th international conference on Information processing in sensor networks (pp. 399-406). ACM.

[12]  D. Raychaudhuri, I. Seskar, M. Ott, S. Ganu, K. Ramachandran, H. Kremo, and M. Singh, (2005, March) "Overview of the ORBIT radio grid testbed for evaluation of next-generation wireless network protocols", In IEEE Wireless Communications and Networking Conference, 2005 (Vol. 3, pp. 1664-1669). IEEE.

[13]  H. Hellbrück, M. Pagel, Köller, A., Bimschas, D., Pfisterer, D., and Fischer, S. (2011, June) "Using and operating wireless sensor network testbeds with WISEBED", In 2011 The 10th IFIP Annual Mediterranean Ad Hoc Networking Workshop (pp. 171-178). IEEE.

[14]  N. BOILLOT, D. DHOUTAUT, J. BOURGEOIS, (2013) "Efficient simulation environment of wireless radio communications in mems modular robots", In : 2013 IEEE International Conference on Green Computing and Communications and IEEE Internet of Things and IEEE Cyber, Physical and Social Computing. IEEE, p. 638-645.

[15]  A. K. Dwivedi, and O. P. Vyas, (2011) "An exploratory study of experimental tools for wireless sensor networks", Wireless Sensor Network, 3(07), 215.

[16]  I. T. DOWNARD, (2004) "Simulating sensor networks in ns-2", NAVAL RESEARCH LAB WASHINGTON DC.

[17]  S. Keshav, (1988) "REAL: A network simulator", Berkeley, Calif, USA: University of California.

[18]  Ns-3 Tutorial. https://www.nsnam.org/docs/release/3.30/tutorial/singlehtml/index.html#futures. [Accessed: 2019-09-25].

[19]  P. S. Katkar, and D. V. R. Ghorpade, (2016) "Comparative study of network simulator: NS2 and NS3", International Journal of Advanced Research in Computer Science and Software Engineering, 6(3).

[20]  P. LEVIS, N. LEE, M. WELSH, C. David, (2003) "TOSSIM: Accurate and scalable simulation of entire TinyOS applications", In : Proceedings of the 1st international conference on Embedded networked sensor systems. ACM, p. 126-137.

[21]  E. PERLA, A. Ó. CATHÁIN, R. S. CARBAJO, M. Huggard, C. Mc Goldrick, (2008, October) "PowerTOSSIM z: realistic energy modelling for wireless sensor network environments", In Proceedings of the 3nd ACM workshop on Performance monitoring and measurement of heterogeneous wireless and wired networks (pp. 35-42). ACM.

[22]  B. L. TITZER, D. K. LEE, J. PALSBERG, "Avrora: Scalable sensor network simulation with precise timing", In : IPSN 2005. Fourth International Symposium on Information Processing in Sensor Networks, IEEE, p. 477-482, 2005.

[23]  UCLA Compilers Group. The AVR Simulation And Analysis Framework. http://compilers.cs.ucla.edu/avrora/, [Accessed: 2019-09-23].

[24]  F. Yu, and R. Jain, (2011) "A survey of wireless sensor network simulation tools", Washington University in St. Louis, Department of Science and Engineering.

[25]  N. Ababneh, (2009, March) "Radio irregularity problem in wireless sensor networks: New experimental results", In 2009 IEEE Sarnoff Symposium (pp. 1-5). IEEE.

[26]  C. H. Wu, and Y. C. Chung, (2007, May) "Heterogeneous wireless sensor network deployment and topology control based on irregular sensor model", In International Conference on Grid and Pervasive Computing (pp. 78-88). Springer, Berlin, Heidelberg.

[27]  "Cgnitive Radio Experimentation World". Internet: http://www.crew-project.eu/wilabt.html, [Jan. 22, 2019].

[28]  S. Bouckaert, W. Vandenberghe, B. Jooris, I. Moerman, and P. Demeester, (2010, May) "The w-iLab.t testbed", In International Conference on Testbeds and Research Infrastructures (pp. 145-154). Springer, Berlin, Heidelberg.

[29]  "USC Networked Systems Laboratory". Internet: https ://nsl.cs.usc.edu/Projects/Tutornet, [Feb. 11, 2018].

[30]  G. WERNER-ALLEN, P. SWIESKOWSKI, M. WELSH, (2005) "Motelab: A wireless sensor network testbed", In : Proceedings of the 4th international symposium on Information processing in sensor networks. IEEE Press, p. 68.

[31]  I. CHATZIGIANNAKIS, S. FISCHER, C. C. KONINIS, G. Mylonas, D. Pfisterer, (2009) "WISEBED: an open large-scale wireless sensor network testbed", In : International Conference on Sensor Applications, Experimentation and Logistics. Springer, Berlin, Heidelberg, p. 68-87.

[32]  "FIT FUTURE INTERNET TESTING FACILITY". Internet: https://fit-equipex.fr/testbeds, [Jan. 5, 2019].

[33]  "NITOS Wireless Sensor Platform". Internet: https://nitlab.inf.uth.gr/NITlab/hardware/sensors/wireless-sensor-platform, [Jan. 28, 2019].

[34]  N. Makris, C. Zarafetas, S. Kechagias, T. Korakis, I. Seskar, and L. Tassiulas, (2015, April) "Enabling open access to LTE network components; the NITOS testbed paradigm", In Proceedings of the 2015 first IEEE Conference on Network Softwarization (NetSoft) (pp. 1-6). IEEE.

[35] "An Open Platform for Developing, Deploying, and Accessing Planetary-Scale services". Internet: https://www.planet-lab.org/, [Jan. 14, 2019].

[36] B. CHUN, D. CULLER, T. ROSCOE, A. Bavier, L. Peterson, M. Wawrzoniak, M. Bowman, (2003) "Planetlab: an overlay testbed for broad-coverage services", ACM SIGCOMM Computer Communication Review, vol. 33, no 3, p. 3-12.

[37] M. OTT, I. SESKAR, R. SIRACCUSA, M. Singh, (2005) "Orbit testbed software architecture: Supporting experiments as a service", In : First International Conference on Testbeds and Research Infrastructures for the DEvelopment of NeTworks and COMmunities, IEEE, p. 136-145.

[38] "Open-Access Research Testbed for Next-Generation Wireless Networks (ORBIT)". Internet: https://www.orbit-lab.org/, [Jan. 26, 2019].

[39] V. Handziski, A. Köpke, A. Willig, and A. Wolisz, (November 2005) "Twist : A scalable and reconfigurable wireless sensor network testbed for indoor deployments", Technical Report TKN-05-008, Telecommunication Networks Group, Technische Universitat Berlin.

[40] "emulab". Internet: https://www.emulab.net/portal/frontpage.php, [ Jan. 02, 2019].

[41] K. Heurtefeux, and F. Valois, (2008, September) "Distributed qualitative localization for wireless sensor networks", In International Conference on Ad-Hoc Networks and Wireless (pp. 218-229). Springer, Berlin, Heidelberg.

[42] E. Niewiadomska-Szynkiewicz, (2012) "Localization in wireless sensor networks: Classification and evaluation of techniques", International Journal of Applied Mathematics and Computer Science, 22(2), 281-297.

[43] "Built With Emulab". Internet: http://www.emulab.arn.dz/index.php/en/pre, [Feb. 02, 2019].

**AUTHORS**

**Souhila Silmi** is currently pursuing her Ph.D. thesis research at the computer science department of USTHB University of Science and Technology Houari Boumedienne, Algiers. She is a member of RIIMA laboratory at USTHB. She has research and teaching experiences. Her research interests include Ad hoc network, wireless sensor networks and Internet of Things.

**Zouina Doukha** received a PHD degree from USTHB University of Science and Technology Houari Boumedienne in 2016. She is associate professor and a member of RIIMA laboratory at USTHB. Her research activities include communication in VANETs, routing, localization and data collection in VSN, WSN and heterogeneous systems. She is also interested in virtualization, SDN and smart cities. She is serving as a reviewer for prestigious international conferences and journals.

**Rebiha Kemcha** received her MSc degree on Networks and Distributed Systems from University of Bejaia, Algeria in 2012. She worked on interconnection networks, their topologies and routing. She is PhD candidate at University of Boumerdes and teaching assistant in security of wireless networks. She works on evolutionary circuit design and design space exploration using multi-objective optimization algorithms.

**Samira Moussaoui** is a full Professor at the computer science department of the University of Sciences and Technologies Houari Boumediene (USTHB), Algiers. She received a PhD degree from USTHB University in 2007. She is Lecturer Researcher in the Department of computer science of USTHB, since 1988. Her research interests include mobile computing systems, mobile and wireless networks, and peer-to-peer systems.